\documentclass[iop,onecolumn]{emulateapj}
\usepackage{natbib,bm,multirow}
\usepackage{graphicx, graphics}
\usepackage{threeparttable}
\pdfoutput=1
\usepackage{amsmath}
\shortauthors{Chung et al.}
\shorttitle{brown dwarf binary}
\newcommand{\te}{t_{\rm E}}

\newcommand{\pie}{\pi_{\rm E}}
\newcommand{\pien}{\pi_{\rm E,N}}
\newcommand{\piee}{\pi_{\rm E,E}}
\newcommand{\bpie}{\bm{\pi}_{\rm E}}
\newcommand{\bmus}{\bm{\mu}_{\rm S}}

\newcommand{\pirel}{\pi_{\rm rel}}
\newcommand{\thetae}{\theta_{\rm E}}
\newcommand{\thetas}{\theta_{\rm \star}}
\newcommand{\cspenalty}{\chi^{2}_{\rm penalty}}
\newcommand{\uas}{\mu{\rm as}}
\newcommand{\murel}{\mu_{\rm rel}}
\newcommand{\dl}{D_{\rm L}}
\newcommand{\ds}{D_{\rm S}}

\newcommand{\delcs}{\Delta \chi^{2}}

\newcommand{\fsspitz}{F_{\rm s,spitzer}}
\newcommand{\fbspitz}{F_{\rm b,spitzer}}

\begin{document}

\title{\textit{Spitzer} Microlensing of MOA-2016-BLG-231L : A Counter-Rotating Brown Dwarf Binary in the Galactic Disk}
\author{Sun-Ju Chung$^{1,2,31}$, Andrew Gould$^{1,3,4,31,33}$, Jan Skowron$^{5,32}$, Ian A. Bond$^{6,34}$, Wei Zhu$^{7,33}$\\
and\\
Michael D. Albrow$^{8}$, Youn Kil Jung$^{1}$, Cheongho Han$^{9}$, Kyu-Ha Hwang$^{1}$, Yoon-Hyun Ryu$^{1}$, In-Gu Shin$^{10}$,\\
 Yossi Shvartzvald$^{11,33}$, Jennifer C. Yee$^{10,33}$, Weicheng Zang$^{12}$, Sang-Mok Cha$^{1,13}$, Dong-Jin Kim$^{1}$, \\
 Hyoun-Woo Kim$^{1}$, Seung-Lee Kim$^{1,2}$, Yun-Hak Kim$^{1,2}$, Chung-Uk Lee$^{1,2}$, Dong-Joo Lee$^{1}$, Yongseok Lee$^{1,13}$, Byeong-Gon Park$^{1,2}$ , Richard W. Pogge$^{3}$\\
(The KMTNet collaboration)\\
Andrzej Udalski$^{5}$, Radek Poleski$^{3,5}$, Przemek Mr{\'o}z$^{5}$, Pawe\l{} Pietrukowicz$^{5}$, Micha\l{} K. Szyma{\'n}ski$^{5}$, \\
Igor Soszy{\'n}ski$^{5}$, Szymon Koz{\l}owski$^{5}$, Krzysztof Ulaczyk$^{5,14}$, Micha\l{} Pawlak$^{5}$\\
(The OGLE collaboration)\\
Charles A. Beichman$^{15}$, Geoffery Bryden$^{16}$, Sebastiano Calchi Novati$^{11}$, Sean Carey$^{17}$, B.\ Scott Gaudi$^{3}$, \\
Calen B. Henderson$^{11}$\\
(The \textit{Spitzer} team)\\
Fumio Abe$^{18}$, Richard Barry$^{19}$, David P. Bennett$^{19,20}$, Aparna Bhattacharya$^{19,20}$, Martin Donachie$^{21}$, \\
Akihiko Fukui$^{22,23}$, Yuki Hirao$^{24}$, Yoshitaka Itow$^{18}$, Kohei Kawasaki$^{24}$, Iona Kondo$^{24}$, Naoki Koshimoto$^{25,26}$, \\
Man Cheung Alex Li$^{21}$, Yutaka Matsubara$^{18}$, Yasushi Muraki$^{18}$, Shota Miyazaki$^{24}$, Masayuki Nagakane$^{24}$, Cl{\'e}ment Ranc$^{19}$, Nicholas J. Rattenbury$^{21}$, Haruno Suematsu$^{24}$, Denis J. Sullivan$^{27}$, Takahiro Sumi$^{24}$, \\
Daisuke Suzuki$^{28}$, Paul J. Tristram$^{29}$, Atsunori Yonehara$^{30}$\\
(The MOA colllaboration)\\
}

\affil{$^1$ Korea Astronomy and Space Science Institute, 776 Daedeokdae-ro, Yuseong-Gu, Daejeon 34055, Korea; sjchung@kasi.re.kr}
\affil{$^2$ Korea University of Science and Technology, 217 Gajeong-ro, Yuseong-gu, Daejeon 34113, Korea}
\affil{$^3$ Department of Astronomy, Ohio State University, 140 W. 18th Ave., Columbus, OH 43210, USA}
\affil{$^4$ Max-Planck-Institute for Astronomy, K{\"o}nigstuhl 17, 69117 Heidelberg, Germany}
\affil{$^5$ Warsaw University Observatory, AI.~Ujazdowskie~4, 00-478~Warszawa, Poland}
\affil{$^6$ Institute of Natural and Mathematical Science, Massey University, Auckland 0745, New Zealand}
\affil{$^7$ Canadian Institute for Theoretical Astrophysics, 60 St George Street, University of Toronto, Toronto, ON M5S 3H8, Canada}
\affil{$^8$ Department of Physics and Astronomy, University of Canterbury, Private Bag 4800 Christchurch, New Zealand}
\affil{$^9$ Department of Physics, Chungbuk National University, Cheongju 361-763, Korea}
\affil{$^{10}$ Harvard-Smithsonian Center for Astrophysics, 60 Garden St., Cambridge, MA 02138, USA}
\affil{$^{11}$ IPAC, Mail Code 100-22, Caltech, 1200 E. California Blvd., Pasadena, CA 91125, USA}
\affil{$^{12}$ Physics Department and Tsinghua Centre for Astrophysics, Tsinghua University, Beijing 100084, People's Republic of China}
\affil{$^{13}$ School of Space Research, Kyung Hee University, Giheung-gu, Yongin, Gyeonggi-do, 17104, Korea}
\affil{$^{14}$ Department of Physics, University of Warwick, Gibbet Hill Road, Coventry, CV4~7AL,~UK}
\affil{$^{15}$ NASA Exoplanet Science Institute, MS 100-22, California Institute of Technology, Pasadena, CA 91125, USA}
\affil{$^{16}$ Jet Propulsion Laboratory, California Institute of Technology, 4800, Oak Grove Dr., Pasadena, CA 91109, USA}
\affil{$^{17}$ Spitzer Science Center, MS 220-6,  California Institute of Technology, Pasadena, CA, USA}
\affil{$^{18}$ Institute for Space-Earth Environmental Research, Nagoya University, Nagoya 464-8601, Japan}
\affil{$^{19}$ Code 667, NASA Goddard Space Flight Center, Greenbelt, MD 20771, USA}
\affil{$^{20}$ Department of Astronomy, University of Maryland, College Park, MD 20742, USA}
\affil{$^{21}$ Department of Physics, University of Auckland, Private Bag 92019, Auckland, New Zealand}
\affil{$^{22}$ Subaru Telescope Okayama Branch Office, National Astronomical Observatory of Japan, NINS, 3037-5 Honjo, Kamogata, Asakuchi, Okayama 719-0232, Japan}
\affil{$^{23}$ Instituto de Astrof\'isica de Canarias, V\'ia L\'actea s/n, E-38205 La Laguna, Tenerife, Spain}
\affil{$^{24}$ Department of Earth and Space Science, Graduate School of Science, Osaka University, Toyonaka, Osaka 560-0043, Japan}
\affil{$^{25}$ Department of Astronomy, Graduate School of Science, The University of Tokyo, 7-3-1 Hongo, Bunkyo-ku, Tokyo 113-0033, Japan}
\affil{$^{26}$ National Astronomical Observatory of Japan, 2-21-1 Osawa, Mitaka, Tokyo 181-8588, Japan}
\affil{$^{27}$ School of Chemical and Physical Science, Victoria University, Wellington, New Zealand}
\affil{$^{28}$ Institute of Space and Astronautical Science, Japan Aerospace Exploration Agency, Kanagawa 252-5210, Japan}
\affil{$^{29}$ University of Canterbury Mt. John Observatory, P.O.Box 56, Lake Tekapo 8770, New Zealand}
\affil{$^{30}$ Department of Physics, Faculty of Science, Kyoto Sangyo University, Kyoto 603-8555, Japan}
\affil{$^{31}$ The KMTNet Collaboration}
\affil{$^{32}$ The OGLE Collaboration}
\affil{$^{33}$ The \textit{Spitzer} Team}
\affil{$^{34}$ The MOA Collaboration}

\begin{abstract}
We analyze the binary microlensing event MOA-2016-BLG-231, which was observed from the ground and from \textit{Spitzer}.
The lens is composed of very low-mass brown dwarfs (BDs) with $M_1 = 21^{+12}_{-5} \ M_J$ and $M_2 = 9^{+5}_{-2}\ M_J$, and it is located in the Galactic disk $D_{\rm L} = 2.85^{+0.88}_{-0.50}\ {\rm kpc}$.
This is the fifth binary brown dwarf discovered by microlensing, and the BD binary is moving counter to the orbital motion of disk stars.
Constraints on the lens physical properties come from late time, non-caustic-crossing features of the \textit{Spitzer} light curve.
Thus, MOA-2016-BLG-231 shows how \textit{Spitzer} plays a crucial role in resolving the nature of BDs in binary BD events with short timescale ($\lesssim 10$ days).
\end{abstract}

\section{introduction}

Brown dwarfs (BDs) are substellar objects that are not massive enough to burn hydrogen.
BDs have a mass between gas giant planets and low-mass stars, and it is thought that the formation and evolution of BDs are different from those of planets and stars \citep{ranc2015}.
Thus, studying BDs is helpful to understand the formation and evolution of stars and planets.

Microlensing is an excellent method to detect faint low-mass objects, such as BDs and planets because it does not depend on the light from the objects, but the mass.
Until now 32 BDs have been detected by microlensing.
Only five of these are isolated BDs, while all the others belong to binary systems.
Microlensing BDs are mostly binary companions to faint M dwarf stars (see Table 1), while most of the many BDs detected by the radial velocity, transit, and direct imaging methods are companions to solar-type stars \citep{ranc2015}.
Hence, microlensing BDs are important to constrain BD formation scenarios including turbulent fragmentation of molecular clouds \citep{boyd&whitworth2005}, fragmentation of unstable accretion disks \citep{stamatellos2007}, ejection of protostars from prestellar cores \citep{reipurth&clarke2001}, and photo-erosion of prestellar cores by nearby very bright stars \citep{whitworth&zinnecker2004}.

However, with microlensing, it is generally difficult to measure the mass of a lens.
This is because we usually obtain only the Einstein timescale,
\begin{equation}
\te \equiv \frac{\thetae}{\murel},
\end{equation}
where $\thetae$ is the angular Einstein radius corresponding to the total lens mass and $\murel$ is the relative lens-source proper motion.
For the measurement of the lens mass, one needs to measure the angular Einstein radius and microlens parallax $\pie$, which yields
\begin{equation}
\label{eqn:mass}
M_{\rm L} = \frac{\thetae}{\kappa \pie}, \qquad
\thetae^{2} = \kappa M_L \pirel,
\end{equation}
where $\pirel \equiv {\rm au}(\dl^{-1} - \ds^{-1})$ is the lens-source relative parallax,  $\dl$ and $\ds$ are the distances to the lens and source, respectively, and $\kappa \equiv 4G/(c^{2}\rm au) \approx 8.14\,{\rm mas}/M_\odot$ \citep{gould2000}.
The angular Einstein radius can be measured from events with finite-source effects, while the microlens parallax can be measured from the detection of light-curve distortions induced by the orbital motion of Earth on a standard microlensing light curve \citep{gould1992,gould2013}.
The large number of microlensing BDs detected to date would appear to indicate that M dwarf-BD binaries and BD binaries are very common.
This is because their mass measurements (hence, unambiguous determination that they are BDs) require clear detection of a microlens parallax signal that can be detected from the ground despite a relatively short timescale.
However, it is usually very difficult to measure the microlens parallax, especially for short timescale events, such as M dwarf-BD binaries.
While the microlensing parallax (derived from Equation(2))\citep{gould2000}
\begin{equation}
\label{eqn:pie}
\pie = \sqrt{\frac{\pirel}{\kappa M_{\rm L}}}
\end{equation}
is on average large for low-mass lenses, this is not by itself usually sufficient to render it measurable in short events.
Rather, large $\pirel$ is required as well.
As a result, half (7/13) of microlensing binaries containing at least one BD and with mass measurements based on ground-based microlensing parallaxes have distances $\dl \lesssim 2\,$kpc, which would be true of a tiny fraction of all microlensing events.
Moreover, of the remainder, all lie in the Galactic disk $\dl <5\,$kpc, and almost all have low, or very low proper motions (so long, or very long timescales), which again is rare.

Hence, it is important to check independently that these relatively frequent detections are not just due to systematics misinterpreted as $``$parallax signal".
The \textit{Spitzer} satellite allows us to do that.
\textit{Spitzer} observations together with ground-based observations yield the microlensing parallax, which does not depend strongly on the event timescale and lens distance.
Thus, \textit{Spitzer} makes it possible to measure the masses of the lenses in short $\te$ binary BD events, which would be quite difficult using only ground-based observations.
The microlens parallax is measured from the difference in the light curves as seen from the two observatories with wide projected separation $D_\perp$ \citep{refsdal1966,gould1994},  which is represented by
\begin{equation}
\bpie \simeq {{\rm au}\over{D_\perp}}\left(\Delta \tau, \Delta \beta_{\pm \pm} \right),  
\end{equation}
where
\begin{equation}
\Delta \tau = {{t_{\rm 0,sat} - t_{\rm 0,\oplus}}\over{\te}};\quad \Delta \beta_{\pm \pm} = \pm u_{\rm 0,sat} - \pm u_{\rm 0,\oplus},
\end{equation}
and where the subscripts indicate the parameters as measured from the satellite and Earth.
Here $t_0$ is the time of the closest source approach to the lens (peak time of the event) and $u_0$ is the separation between the lens and the source at time $t_0$.

In this paper, we report the discovery of the fifth binary BD from the analysis of the microlensing event  MOA-2016-BLG-231, which was observed from the ground and from \textit{Spitzer}.
Although \textit{Spitzer} can identify binary BDs events with short $\te$, observing such events is extremely challenging because of the short timescale and the $3-9$ day observation delay (see Figure 1 of \citealt{udalski2015}).
Thus, \textit{Spitzer} does not usually observe caustic crossings of such events.
However, we here show that even non caustic-crossing \textit{Spitzer} light curves can resolve the nature of a binary BD lens.

\section{OBSERVATIONS}
\subsection{Ground-based observations}

The microlensing event MOA-2016-BLG-231 was first alerted on UT 22:18 6 May by the Microlensing Observations in Astrophysics (MOA; \citealt{suzuki2016}). MOA uses a 1.8 m telescope with 2.2 deg$^2$ field-of-view (FOV) at Mt. John Observatory in New Zealand.
The lensed source star is at $(\alpha, \delta)_{\rm J2000}$ = $(17^{\rm h}53^{\rm m}12^{\rm s}.0,-30^{\circ}11'32\farcs1)$, corresponding to $(l, b) = (359 \fdg 77,-2 \fdg 06)$.
The Early Warning System (EWS) of the Optical Gravitational Lensing Experiment (OGLE) collaboration \citep{udalski2003} also alerted this event 7 days after the MOA alert.
OGLE uses the 1.3 m Warsaw telescope at the Las Campanas Observatory in Chile.
The event is located in the OGLE field BLG501, which is observed with cadence $\Gamma \simeq 0.4\,{\rm hr^{-1}}$.
The event is designated as OGLE-2016-BLG-0864 by OGLE.
Here we note that although MOA first alerted the event, \textit{Spitzer} observations were triggered by OGLE data rather than MOA, and MOA did not play a major role in characterization of the lens.
Since the Einstein crossing time of the event is short $\te \sim 13\ \rm days$, and the OGLE baseline is slightly variable on long timescales, we used only 2016 season data sets of OGLE and MOA for light curve modeling.

The event was also observed by the Korea Microlensing Telescope Network (KMTNet; \citealt{kim2016}).
KMTNet uses 1.6 m telescopes with 4.0 deg$^2$ FOV at CTIO in Chile (KMTC), SAAO in South Africa (KMTS), and SSO in Australia (KMTA).
MOA-2016-BLG-231 lies in two overlapping KMTNet fields, BLG01 and BLG42, with a combined cadence of $\Gamma=4\,{\rm hr}^{-1}$.
It is designed by KMTNet as KMT-2016-BLG-0285 \citep{kim2018}.
Most of KMTNet data were taken in $I$ band, and for the characterization of the source star, some data were taken in $V$ band from CTIO.
The KMTNet data were reduced by pySIS based on the difference imaging method \citep{alard1998, albrow2009}. 

\subsection{\textit{Spitzer} observations}

Since 2014, \textit{Spitzer} has been observing microlensing events toward the Galactic bulge in order to measure the microlens parallax.
While the overwhelming majority of events chosen for {\it Spitzer} observations are (apparently) generated by point-mass lenses at the time of selection, in accordance with the detailed protocols described by \citet{yee2015}, the {\it Spitzer} team does select known binary and planetary lenses whenever there appears to be a reasonable chance to measure the microlens parallax.  
MOA-2016-BLG-231 was such a case.  
At the time of selection, the event was regarded by the team as ``difficult, but maybe doable'' because it was recognized that the timescale of event was quite short and the first {\it Spitzer} observation would be 15 days after the final anomalous feature in the light curve. 
See Figure~1.
It was observed for three weeks, mostly at a cadence of $\sim 1\,{\rm day}^{-1}$, but roughly double that for the first few days because of the limited number of available events due to {\it Spitzer}'s sun-angle exclusion of more easterly targets.

\section{LIGHT CURVE ANALYSIS}
\label{sec:lc-anal}
\subsection{Ground-based data}

The observed ground-based data of the event MOA-2016-BLG-231 have a clear caustic-crossing feature, while the \textit{Spitzer} data (as anticipated) show only a general decline.
We therefore begin by incorporating only ground-based data to conduct binary lens modeling.
Standard binary lens modeling requires seven parameters including three single-lens parameters ($t_0$, $u_0$, $\te$) and four additional parameters: the projected separation of the lens components in units of $\thetae$ ($s$), the mass ratio of the components ($q$), the angle between the source trajectory and the binary axis ($\alpha$), and the normalized source radius ($\rho = \thetas/\thetae$) \citep{rhie1999}, where $\thetas$ is the angular radius of the source.
In addition, there are two flux parameters for each observatory, the source flux $f_{s,i}$ and blended flux $f_{b,i}$ of the \textit{i}th observatory.
The two flux parameters at a given time $t_j$ are modeled by
\begin{equation}
F_i(t_j) = f_{s,i}A_{i}(t_j) + f_{b,i},
\label{eqn:ftot}
\end{equation}
where $A_i$ is the magnification as a function of time at the \textit{i}th observatory \citep{rhie1999}.
The two flux parameters of each observatory are determined from a linear fit.

We conduct a grid search in the parameter space $(s, q, \alpha)$ to find the best-fit model.
The ranges of the parameters are $-1 \leqslant {\rm log} s \leqslant 1$, $-2 \leqslant {\rm  log} q \leqslant 0 $, and $0 \leqslant \alpha \leqslant 2\pi$, respectively.
During the grid search, the other parameters are searched for using by a Markov Chain Monte Carlo (MCMC) method.
The magnification is calculated by inverse ray shooting near and in the caustic \citep{kayser1986, schneider1988, wambsganss1997} and multipole approximations \citep{pejcha2009,gould2008} otherwise.
From this, we find only one local minimum at $(s, q, \alpha) \simeq (1.3, 0.4, 4.2)$.
We then seed the local solutions into the MCMC for which all parameters are allowed to vary, and finally find a global solution of the binary lens model.

As in many binary and planetary events, $\rho$ can be measured from the effect of the finite size of the source to smooth the intrinsically divergent magnification profile of the caustic.
Because the source crosses the caustic, we consider the limb-darkening variation of the finite source star in the modeling.
For this, we adopt the source brightness profile, which is approximated by
\begin{equation}
S_{\lambda} = {F_{\lambda}\over{\pi \theta^{2}_{\star}}}\left[1 - \Gamma_{\lambda}\left(1 - {3\over{2}}\cos\phi\right)\right],
\end{equation}
where $F_{\lambda}$ is the total flux of the source at wavelength $\lambda$, $\Gamma_{\lambda}$ is the limb darkening coefficient, and $\phi$ is the angle between the normal to the surface of the source star and the line of sight \citep{an2002}.
According to the source type, which is discussed in Section 4, we adopt that $\Gamma_{I} = 0.54$ , $\Gamma_{V} = 0.711$, and $\Gamma_{L} = 0.178$ from \citet{claret2000} and \citet{claret2011}.

\subsection{Combination of ground-based and \textit{Spitzer} data}

Thanks to the \textit{Spitzer} data, we can constrain the higher-order effects of microlensing parallax and lens orbital motion, even though $\te \simeq 13$ days would not be long enough to detect these two effects using ground-based data alone.
When including the parallax effect in the model, it is important to include also the orbital motion effect because of the degeneracy between the two \citep{skowron2011, batista2011, han2016}.
Thus, we conduct the modeling with both parallax and orbital effects.
The microlens parallax enters as a two-parameter vector $\bpie = (\pien,\piee)$, whose amplitude is given by Equation (4) and whose direction is that of the lens-source relative proper motion in the geocentric frame, i.e., $\bpie = \pie(\bm{\mu}/\mu)$.
Under the approximation of linear orbital motion of the binary lens, the orbital motion effect is described by two parameters, $ds/dt$ and $d\alpha/dt$, which are the change rates of the binary separation and the orientation angle of the binary axis, respectively.
Hence, four additional parameters are added in the model.
In contrast to the ground-based light curve, \textit{Spitzer} covers only the falling wing of the light curve.
Thus, it is essential to incorporate the color constraint between OGLE and \textit{Spitzer} $(I_{\rm ogle} - L)$ in order to find the correct parallax solution. See, for example, the analysis of OGLE-2016-BLG-0168 by \citet{shin2017}.
We find $(I_{\rm ogle} -L) = 1.911 \pm 0.020$ by combining the source instrumental $(V-I)$ color measurement, which is discussed in Section 4, with a $VIL$ instrumental color-color relation derived from matched field stars.
To enforce the color constraint, we add a $\cspenalty$ to $\chi^{2}$ calculated in the model, which is defined as
\begin{equation}
\cspenalty = {\lbrace(I_{\rm ogle} - L) + 2.5{\rm log}(F_{\rm s, ogle}/F_{\rm s,spitzer})\rbrace^{2}\over{\sigma_{c}^{2}}},
\end{equation}
where $F_{\rm s, ogle}$ and $F_{\rm s, spitzer}$ are the source fluxes of OGLE and \textit{Spitzer}, which are obtained from the model,  and $\sigma_c$ is the error of the color constraint $(I_{\rm ogle} -L)$.
Thus, $\chi^{2}_{\rm total}$ = $\chi^2 + \cspenalty$.
In order to find correct source and blended fluxes of \textit{Spitzer},  $F_{s,spitzer}$ and $F_{b,spitzer}$, with a strong color constraint, we include them as chain variables when modeling.
Because of the low value of $\te$, we expect most of the microlens parallax ``signal" to come from \textit{Spitzer} and not the ground-based data.
However, we find that when we model the event using all data sets, most of these contribute signal at the $\delcs$ of few tens level (and with different signs), with these signals coming overwhelmingly from the wings of the event, as determined from a cumulative $\delcs$ plot (see further below).
Such false parallax signals are not uncommon in MOA data, and have also been seen in KMTNet data during its much shorter history.
Therefore, we restrict all ground-based data sets except OGLE to the time interval $7530.0 < {\rm HJD}^\prime  < 7544.0$, where the rapid changes in magnification ensure that very low-level systematics will not play any significant role.
We then find that the cumulative distribution of $\delcs = \chi^2_{\rm parallax+orbital}-\chi^2_{\rm standard}$ shows no strong trends in any of the data sets except \textit{Spitzer}.
See Figure 2.

Observations from the ground and from \textit{Spitzer} yield a well-known four degeneracy for the microlens parallax : $(+,+)$, $(+,-)$, $(-,+)$, and $(-,-)$, which register the signs of the impact parameters as measured from the ground and \textit{Spitzer}, respectively \citep{zhu2015}.
From the modeling, we find that the event MOA-2016-BLG-231 has only two solutions, $(+,-)$ and $(-,-)$, and that the $(+,-)$ solution is preferred by $\delcs = 1.92$.
The other models $(+,+)$ and $(-,+)$ converge to $(+,-)$ and $(-,-)$ models, respectively.
The best-fit lensing parameters of the $(+,-)$ and $(-,-)$ models are presented in Table 2.
Figure 1 shows the best-fit light curve of the event, i.e., for the $(+,-)$ model.
The corresponding source trajectories for the ground and \textit{Spitzer} are presented in Figure 3.
Figure 3 also shows the caustic structures and positions of the lens components at three different epochs, $t_1 = 7531.9$ (caustic entrance), $t_2 = 7537.3$ (caustic exit), and $t_3 = 7570.0$ (close to baseline).
As shown in Figure 3, the caustics and lens positions at the two epochs ($t_1$ and $t_2$) appear almost the same.
This is because the characteristic orbital timescale, $\gamma^{-1}\simeq 1.0\,{\rm yr}$ is long compared to the time interval explored $t_2-t_1\simeq 5\,{\rm days}$.
Here, $\bm{\gamma} \equiv (\gamma_\parallel,\gamma_\perp) = (ds/dt/s,d\alpha/dt)$.
While the $\chi^2$ improvement of the parallax+orbital solution compared to the parallax-only solution is relatively small, $\Delta\chi^2 \simeq 9$, we will argue in Section 5 that this detection of orbital motion is likely real.

Figure 4 shows $\delcs$ distributions of the parallax and orbital motion parameters for the $(+,-)$ and $(-,-)$ models.
The parallax amplitude $\pie=|\bm{\pie}|$ is quite well constrained and very similar in the two cases, implying that the mass and distance of the system will be both well measured and not seriously impacted by the two-fold parallax degeneracy.
From this, we find that even though \textit{Spitzer} has no caustic-crossing features and it has only fragmentary coverage of the light curve, we can constrain the physical properties of lens.

In Figure 4, we mark four representative models for $(+,-)$, which are located just inside the $3\sigma$ contour.
The corresponding parallax and orbital parameters are presented in Table 2.
Figure 5 shows the \textit{Spitzer} trajectories and resulting light curves of the four models.
As shown in Figure 5, these four models have different trajectories from the best-fit one, and hence dramatically different predicted {\it Spitzer} light curves over the peak of the event. 
However, during the time interval that {\it Spitzer} actually took data, all four predict similar light curves (see the inset to Figure 5). 
Despite these different trajectories, and as discussed above, these have qualitatively similar amplitudes, $\pie$, which is what enables a mass measurement.

\section{Estimate of $\thetae$}

As mentioned in Section 1, $\thetae$ and $\pie$ should be measured for the measurements of the mass and distance of the lens.
Thanks to caustic-crossing features, one measure $\rho$ from the modeling, while $\thetae = \thetas/\rho$.
Thus, we need to estimate $\thetas$ for the measurement of $\thetae$.
We estimate $\thetas$ from the intrinsic color and brightness of the source, which can be derived by the offset between the source and the red clump in the instrumental color and magnitude diagram (CMD) (see Fig. 6) \citep{yoo2004}.
These are determined by
\begin{equation}
[(V-I),I]_{\rm s,0} = [(V-I),I]_{\rm clump,0} + [\Delta(V - I), \Delta I],
\end{equation}
where $ [\Delta(V - I), \Delta I]$ is the offsets of the color and brightness between the source and the clump.
We determine the instrumental source color by regression of the $V$ on $I$ flux measurements and derive the source instrumental magnitude
from the model.
These have errors of 0.004 and $0.020$ mag, respectively.
We then centroid the clump in color and magnitude with errors of 0.022 and 0.05 mag respectively.
The measured color and magnitude offsets are $ [\Delta(V - I), \Delta I] =[0.06 \pm 0.02, -0.19 \pm 0.05]$.
Adopting $[(V-I), I]_{\rm clump,0} = (1.06,14.44))$ from \citet{bensby2011} and \citet{nataf2013}, we find $[(V - I), I]_{\rm s,0} = [1.12, 14.25]$.
This indicates that the source is a K-type giant.
We determine the angular radius of the source $\thetas$ using $VIK$ color-color relation \citep{bessell1988} and the color/surface brightness relation \citep{kervella2004}.
As a result, we find that $\thetas = 7.23 \pm 0.40\ \uas$.
With the measured $\rho$ and $\thetas$, we determine the angular radius of the Einstein ring corresponding to the total mass of the lens,
\begin{equation}
\thetae = \thetas/\rho = 0.233^{+0.013}_{-0.014}\ \textrm{mas}.
\end{equation}
The relative lens-source proper motion is
\begin{equation}
\murel = \thetae/\te = 6.57\pm 0.37\ \textrm{mas\ yr$^{-1}$}.
\end{equation}

\section{Lens properties}

Using the estimated $\thetae$ and $\pie = 0.99^{+0.31}_{-0.35}$, we measure the total mass of the lens system,
\begin{displaymath}
M = {\thetae\over{\kappa \pie}} = 0.029^{+0.016}_{-0.007}\ M_\odot.
\end{displaymath}
The lens is composed of low-mass BDs with masses $M_1 = 0.020^{+0.011}_{-0.005}\ M_\odot$ and $M_2 = 0.009^{+0.005}_{-0.002}\ M_\odot$, where $M = M_1 + M_2$, and the projected separation of the two BDs is $a_\perp = 0.88^{+0.27}_{-0.15}$ au.
The relative parallax between the lens and the source is
\begin{equation}
\pirel = \thetae \pie = 0.23^{+0.07}_{-0.08}\ {\rm mas}.
\end{equation}
Assuming that the source is located at 8.3 kpc \citep{nataf2013}, we estimate the distance to the lens,
\begin{equation}
D_{\rm L} = \left({\pirel\over{\rm au}} + {1\over{D_{\rm S}}}\right)^{-1} = 2.85^{+0.88}_{-0.50}\ {\rm kpc}.
\end{equation}
Hence, the lens is a BD binary located in the Galactic disk.
This is the fourth BD binary discovered by microlensing.
Because $\thetae$ is well measured due to a precise $\rho$ measurement, which comes from good coverage of caustic-crossing, the errors in the mass and distance of the lens primarily reflect the error in $\pie$.
As shown in Figure 4, the correlation between $\pien$ and $\piee$ components is not well approximated by a linear relation.
Hence, we use the best-fit MCMC chains to determine the errors in physical lens parameters including the mass and distance of the lens.
Then, one can determine the standard deviation of each physical parameter from the chains.
Thus, physical lens parameters in Table 3 represent the median values of each physical parameter from the $(+,-)$ and $(-,-)$ MCMC chains, and their error bars  represent the 16th and 84th percentile values of each parameter from the chains.

In order to check that the binary lens is a bound system, we compute $\beta$, i.e., the ratio of the projected kinetic to potential energy \citep{an2002},
\begin{equation}
\beta \equiv \left({\rm KE\over{\rm PE}}\right)_\perp = {{(s\thetae D_{\rm L}/{\rm au}})^3(\gamma^2\ {\rm yr}^2)\over{M/M_\odot}}  = 0.16^{+0.20}_{-0.10}
\end{equation}
where $\gamma = [(ds/dt/s)^2 + (d\alpha/dt)^2]^{1/2}$.
Since $\beta=0.16^{+0.20}_{-0.10}$ (or $\beta=0.52^{+0.30}_{-0.21}$ for $(-,-)$ model) represents very typical values for a bound pair seen at random orientation, and in particular indicates that the lens system satisfies the condition of a bound system, $\beta < 1$ \citep{an2002}, it is valid.
We list estimated physical parameters of the lens system in Table 3.
In Table 3, we also list physical lens parameters for four models with $\delcs \sim 9$ from Figures 4 $\&$ 5.
Three of the four models imply that the lens system is a low-mass BD binary in the disk, similar to the best-fit solutions, while for model 1 it is a binary composed of a low-mass star and a BD in the disk.
This is because of large uncertainties of $\pien$, as shown in Figure 4.
We further discuss these models of the lens system in Section 6.

\section{DISCUSSION}
Table 2 shows that all $\piee$ for two best models is negative.
This means that the lens located in the Galactic disk is moving in the opposite direction to the disk orbital motion.
It is unusual.
In order to check that it is real, we conduct the modeling under the condition that high-order effect parameters set to zero  and the initial values of the other standard model parameters including \textit{Spitzer} fluxes set to the best-fit solutions of the two parallax+orbital models including $(+,-)$ and $(-,-)$.
We also conduct the same test for the four models indicated in Figure 4.
As a result, we find that for all of the six models the slope of the \textit{Spitzer} fluxes is steeper than the model (see Figure 7).
This means that \textit{Spitzer} observed the event later than Earth, and the lens is moving toward \textit{Spitzer}, i.e., toward the west.
Therefore, it is real that the lens is moving in the opposite direction to the disk orbital motion.
In addition, in order to  demonstrate that the lens is moving west relative to the source, we measure the proper motion of the source (see Figure 8).
We find that $\bmus = (\mu_{\rm S,N}, \mu_{\rm S, E}) = (0.021 \pm 0.699, 0.231\pm0.699)\ {\rm mas\ yr^{-1}}$.
As shown in Figure 8, the source is certainly part of the bulge population and is not moving relative to the bulge stars.
Hence, the measurement of the lens-source relative motion clearly means that the lens is not moving with the disk.
With the measurement of the source proper motion, we can measure the heliocentric proper motion of the lens,
\begin{equation}
\bm{\mu}_{\rm L,hel} = \bmus + \bm{\mu}_{\rm rel} + {\pirel\over\rm au}\bm{v}_{\oplus,\perp},
\end{equation}
where $\bm{v}_{\oplus,\perp}$ is the velocity of Earth at the peak of the event and projected perpendicular to the directory of the event.
For this case, $\bm{v}_{\oplus,\perp} = (v_{N}, v_{E}) = (1.09, 26.89)\ \rm km\ s^{-1}$.
From Equation (14), we find that  $(\mu_{\rm L,hel,N}, \mu_{\rm L,hel,E}) = (-6.16^{+0.40}_{-0.38},-0.37^{+0.25}_{-0.39})\ {\rm mas\ yr^{-1}}$ for  $(+,-)$ solution, while for  $(-,-)$ solution, $(\mu_{\rm L,hel,N}, \mu_{\rm L,hel,E}) = (-6.09^{+0.43}_{-0.39},-0.87^{+0.34}_{-0.74})\ {\rm mas\ yr^{-1}}$.
From the measurement of $\bm{\mu}_{\rm L,hel}$, we can also measure the transverse velocity of the lens, \citep{shvartzvald2018},
\begin{equation}
\bm{v}_{\rm L,\perp} =  \dl\bm{\mu}_{\rm L,hel} + (1-{\dl\over{\ds}})\bm{v}_\odot,
\end{equation}
where $\bm{v}_\odot = \bm{v}_{\odot,\rm pec} + \bm{v}_{\odot,\rm cir}$.
Here $v_{\odot,\rm pec}(l,b) = (12,7)\ \rm km\ s^{-1}$ are the transverse components of the Sun's peculiar velocity relative to the Local Standard of Rest and $v_{\odot,\rm cir}(l,b) = (220,0)\ \rm km\ s^{-1}$ is the disk circular velocity.
Thus, the peculiar velocity of the lens relative to the mean motion of the Galactic disk stars \citep{shvartzvald2018} is
\begin{eqnarray}
\bm{v}_{\rm L, pec} & =& \bm{v}_{\rm L,\perp} - \bm{v}_{\rm L,\rm cir} \\
 &=& \dl\bm{\mu}_{\rm L,hel} - \bm{v}_{\odot,\rm cir}{\dl\over{\ds}} + \bm{v}_{\odot,\rm pec}\left(1-{\dl\over{\ds}}\right),\nonumber
\end{eqnarray}
where $\bm{v}_{\rm L,cir} = \bm{v}_{\odot,\rm cir}$ because both lens and Sun are disk stars.
From Equation (16), we find that for two degenerate solutions, $(+,-)$ and $(-,-)$, $v_{\rm L,pec}(l,b) = (-144.12^{+37.69}_{-50.40}, -27.78^{+5.95}_{-7.79})\ \rm km\ s^{-1}$ and $v_{\rm L,pec}(l,b) = (-172.29^{+38.72}_{-66.15}, -24.98^{+7.17}_{-7.71})\ \rm km\ s^{-1}$, respectively.
This means that the lens is counter-rotating relative to the motion of Galactic disk stars.
In order to check that the four models with $\delcs \sim 9$ in Figure 4 are compatible with the two solutions, we also estimate the proper motions and peculiar velocities of the lens for the four models, and they are as follows as
\begin{displaymath} 
\mu_{\rm L,hel}(N,E) = \left\lbrace 
\begin{array}{llll}
$(-3.67,-5.03)$\ \rm{mas\ yr^{-1}} & \textrm{for model 1} \\
$(-6.35,-0.03)$\ \rm{mas\ yr^{-1}} & \textrm{for model 2}  \\
$(-6.01,-0.72)$\ \rm{mas\ yr^{-1}} & \textrm{for model 3} \\
$(-5.87,-0.10)$\ \rm{mas\ yr^{-1}} & \textrm{for model 4},  \\
\end{array}\right.
\end{displaymath}

\begin{displaymath} 
v_{\rm L,pec}(l,b) = \left\lbrace 
\begin{array}{llll}
$(-331.79, 85.82)$\ \rm{km\ s^{-1}} & \textrm{for model 1} \\
$(-160.13,-38.37)$\ \rm{km\ s^{-1}} & \textrm{for model 2}  \\
$(-129.25,-20.73)$\ \rm{km\ s^{-1}} & \textrm{for model 3} \\
$(-90.08, -18.29)$\ \rm{km\ s^{-1}} & \textrm{for model 4}.  \\
\end{array}\right.
\end{displaymath}
The lens proper motions of the four models are consistent with those of the two best-fit solutions.
Therefore, the lens is a counter-rotating disk object.
Until now two counter-rotating disk objects (OGLE-2016-BLG-1195L \citep{shvartzvald2017}, OGLE-2017-BLG-0896L \citep{shvartzvald2018}) have been discovered by microlensing, the first being right at the hydrogen-burning limit and the second being a low-mass BD.
MOA-2016-BLG-231L is the third such object, and it is the first counter-rotating BD binary discovered by \textit{Spitzer}.
The unusual kinematics of the BD binary suggest that the BD binary could be a halo object or a member of counter-rotating low-mass object population, as mentioned in \citet{shvartzvald2018}.

For the event MOA-2016-BLG-231, it is found that there is an extremely small offset between the source position and the baseline object.
The offset is 0.05 pixels, and it corresponds to $0.05\times0.26\ \rm arcsec = 13$ mas.
This means that the blend could be associated with the event.
However, the blended flux is in fact consistent with zero.
First, the formal estimate of the blended flux is $f_b=0.33 \pm 0.17$, where one unit of flux corresponds to $I=18$.
This in itself is consistent with zero at the $2\,\sigma$ level.
Moreover, the estimate of $f_b$ is ultimately derived from $f_b=f_{\rm base}-f_s$, where $f_s$ is the source flux from the microlensing model and $f_{\rm base}$ comes from DoPhot \citep{schechter1993} photometry of this field location.
Because of the mottled background of unresolved turnoff stars in these croweded fields, $f_{\rm base}$ can easily have errors at this level.
Therefore, there is no clear evidence for blended light.
In addition, we measure $I_{\rm base}$ using OGLE deep stack images.
From this, we find that $I_{\rm base}$ from the deep stack images is 0.06 mag brighter than the value obtained from the OGLE reference images (i.e. normal OGLE data) due to a nearby star at $0.6''$.
Hence, $I_{\rm base}$ cannot be estimated from the reference image photometry to better than 0.06 mag due to the presence of the nearby star.
Therefore blended light cannot be determined better than 0.5 flux units.
This reinforces the naive conclusion that there is no evidence for light from the lens.

The result of modeling with chain variables $\fsspitz$ and $\fbspitz$ yields negative $\fbspitz$ (see Table 2).
$\fbspitz \sim -4.5$ (see Table 2).
This is somewhat unusual.
As reported in \citet{calchinovati2015} and \citet{shvartzvald2018}, the negative \textit{Spitzer} blending can be generated when the flux of unresolved faint stars is included in the global background flux.
The event OGLE-2017-BLG-0896 \citep{shvartzvald2018} is also the event affected by the excess flux due to unresolved stars and has almost the same \textit{Spitzer} blending as this event.

The result of this study indicates that the lens system of MOA-2016-BLG-231 consists of two low-mass BDs.
In principle, there are two scenarios that would drive us to very different conclusions, but neither of them is likely to be true.
First, a smaller parallax value would give a more massive and more distant lens system.
For example, out of the four models indicated in Figure 4, model 1 has the smallest parallax, which produces the most massive and the most distant lens system and thus might in principle solve the puzzle on the counter-rotating motion.
However, the model 1 gives a low-mass M dwarf-BD binary with $M_{\rm tot}=0.18\ M_\odot$ at a distance of $\dl = 6.3\ \rm kpc$, which still implies that it is in the disk.
Moreover, with proper motion $\mu_{\rm L,hel}(N,E)= (-3.7,-5.0)$, it is still counter-rotating.
Therefore, the smaller parallax scenario is unlikely to resolve the issue of a counter-rotating lens.
Furthermore, for the other models, all the lens systems are low-mass BD binaries with $M_{\rm tot} = 0.02 \sim 0.03\ M_\odot$ that is located at the disk $\dl < 3.2\ \rm kpc$ and their proper motions are consistent with the two best-fit models (see Table 3), as mentioned before.

Another possibility could be that such an unusual solution arises from the unknown systematics in the \textit{Spitzer} photometry. 
\citet{poleski2016} and \citet{zhu2017} have observed systematics on timescales of tens of days.
In principle, the \textit{Spitzer} light curve could be affected by such systematics, but they cannot be recognized because the total duration of the light curve is short. However, \citet{zhu2017}, based on the \textit{Spitzer} event sample that was uniformly analyzed in that work, concluded that such long-term systematic trends in the \textit{Spitzer} photometry appear in $< 5\%$ of all cases.
Therefore, there is a $< 5\%$ probability that systematics in the \textit{Spitzer} photometry led to our current solution.
The proper motion is $6.6\ \rm mas\ yr^{-1}$ (independent of the parallax measurement).
Thus, the solution can be checked with adaptive optics observations at first light of next-generation, thirty-meter class telescopes.
For example, in 2028 the source and lens will be separated by $79\ \rm mas$, which is $\sim 6$ times the FWHM at 1.6 microns for a 30-m telescope, so easily resolved even though the source is a giant.
If the lens really is a brown dwarf, no light should be detected.
However, if the solution is wrong (e.g. due to unknown systematics), then the lens should be more massive, i.e. a star.
In that case, light from the lens should be directly detected.
Furthermore, if light is detected, this will also allow a check of the direction of the source-lens relative proper motion.

\section{CONCLUSION}

We present the analysis of the binary lensing event MOA-2016-BLG-231 that was observed from the ground and from \textit{Spitzer}.
Even though \textit{Spitzer} did not cover caustic-crossing parts and it covered only partial wing parts of the light curve, we could determine the physical properties of the lens.
It is found that the lens is a binary system composed of low-mass BDs with $M_1 = 21^{+12}_{-5}\ M_J$ and $M_2 = 9^{+5}_{-2}\ M_J$, and it is located in the Galactic disk $D_{\rm L} = 2.85^{+0.88}_{-0.50}\ {\rm kpc}$.
The BD binary is moving counter to the orbital motion of disk stars. 
This solution can be checked in the future with adaptive optics observations with thirty-meter class telescopes.
This result shows how \textit{Spitzer} plays a crucial role in short $\te$ BD events, despite the fact that it covered only the declining wing of the light curve in a relatively short event.

\acknowledgments
Work by S.-J. Chung was supported by the KASI (Korea Astronomy and Space Science Institute) grant 2018-1-830-02.
Work by WZ, and AG were supported by AST-1516842 from the US NSF.
WZ, IGS, and AG were supported by JPL grant 1500811.
Work by C.H. was supported by grant 2017R1A4A1015178 of National Research Foundation of Korea.
Work by C.R. was supported by an appointment to the NASA Postdoctoral Program at the Goddard Space Flight Center, administered by USRA through a contract with NASA.
This research has made use of the KMTNet system operated by KASI and the data were obtained at three host sites of CTIO in Chile, SAAO in South Africa, and SSO in Australia.
The MOA project is supported by JSPS KAKENHI Grant Number JSPS24253004, JSPS26247023, JSPS23340064, JSPS15H00781, JP16H06287, and JP17H02871.
The OGLE has received funding from the National Science Centre, Poland, grant MAESTRO 2014/14/A/ST9/00121 to A.U.


\clearpage
\begin{deluxetable}{lccccccc}
\tabletypesize{\scriptsize}
\tablewidth{0pt}
\tablecaption{Microlensing brown dwarfs in binaries. \label{tbl-one}}
\tablehead{
 \colhead{Event}   & \colhead{$M_{\rm host}$}  & \colhead{$M_{\rm comp}$} & \colhead{$D_{\rm L}$}  & \colhead{$\te$} & \colhead{$\mu$} & \colhead{${\pi_{\rm E}} ^{\dagger}$}   & \colhead{Reference}\\
 \colhead{ }          &   \colhead{($M_{\odot}$)} & \colhead{($M_{\rm J}$)} &   \colhead{( kpc)}            &  \colhead{(days)}   &  \colhead{(mas/yr)} & \colhead{ } & \colhead{ }
 }
 \startdata
OGLE-2006-BLG-277                       &  $0.10 \pm 0.03$   & $52 \pm 15$       & $0.60 \pm 0.14$  & $37.9 \pm 0.1$     &  $13.0 \pm 1.1$  & $1.13 \pm 0.16$  & (1) \\
MOA-2007-BLG-197$^{\rm (a)}$   &  $0.82 \pm 0.04$   & $41 \pm 2$         & $4.17 \pm 0.30$  & $82.3 \pm 1.2$    &  $4.0 \pm 0.2$      & $-$                       & (2) \\
OGLE-2009-BLG-151                        &  $0.018 \pm 0.001$ & $7.9 \pm 0.3$  & $0.39 \pm 0.01$  & $28.0 \pm 0.1$   &  $9.3 \pm 0.1$     &  $3.45$                   & (3) \\
MOA-2010-BLG-073                       &  $0.16 \pm 0.03$   & $11.0 \pm 2.0$   & $2.8 \pm 0.4$      & $44.3 \pm 0.1$     &  $4.6 \pm 0.4$    & $0.37$                    & (4) \\
OGLE-2011-BLG-0420                     &  $0.025 \pm 0.001$   & $9.9 \pm 0.5$ & $1.99\pm 0.08$   & $35.2 \pm 0.1$     &  $3.4 \pm 0.3$    & $1.17$                     & (3) \\
MOA-2011-BLG-104                          &   $0.18 \pm 0.11$     & $21 \pm 10$       & $3.29 \pm 1.20$   &  $39.3 \pm 0.5$   & $5.0 \pm 0.7$     &  $0.34 \pm 0.21$   & (5) \\
MOA-2011-BLG-149                         &  $0.14 \pm 0.02$   & $20 \pm 2$         & $1.07 \pm 0.10$  & $179.7 \pm 8.5$   & $2.1 \pm 0.2$     & $0.78 \pm 0.04$      & (5) \\
OGLE-2012-BLG-0358                     &  $0.022 \pm 0.002$   & $1.9 \pm 0.2$     & $1.73 \pm 0.12$  & $26.5 \pm 0.1$     &  $4.0 \pm 0.4$    & $1.50$                & (6) \\
OGLE-2013-BLG-0102                     &  $0.10 \pm 0.01$   & $12.6 \pm 2.1$   & $3.04 \pm 0.31$  & $37.6 \pm 0.4$     &  $4.2 \pm 0.4$    & $0.48$                    & (7) \\
OGLE-2013-BLG-0578                     &  $0.12 \pm 0.01$   & $33.5 \pm 4.2$   & $1.16 \pm 0.11$  & $72.1 \pm 0.8$     &  $4.9 \pm 0.4$    & $0.77$                      & (8) \\
OGLE-2014-BLG-0257                     &  $0.19 \pm 0.02$   & $37.7 \pm 5.2$   & $1.25 \pm 0.13$  & $77.9 \pm 1.4$     &  $5.3 \pm 0.4$    & $0.60$                    & (9) \\
OGLE-2014-BLG-1112                       &  $1.07 \pm 0.28$   & $31.8 \pm 8.2$   & $4.84 \pm 0.67$  & $106.4 \pm 0.9$   &  $3.2 \pm 0.5$    & $0.08$                     & (10) \\
OGLE-2015-BLG-1319$^{\rm (b)}$  &  $0.60 \pm 0.07$   & $59.1 \pm 4.1$   & $4.84 \pm 0.13$  & $98.8 \pm 4.7$     &  $2.4 \pm 0.2$    & $0.12$                     & (11) \\
OGLE-2016-BLG-0693                     &  $0.86 \pm 0.24$   & $42.9 \pm 16.5$ & $4.85 \pm 0.67$  & $142.3 \pm 49.9$ &  $1.7 \pm 0.4$    & $0.09$                  & (12) \\
OGLE-2016-BLG-1195$^{\rm (c)}$  &  $0.08 \pm 0.01$   & $0.004 \pm 0.001$  & $3.91 \pm 0.42$ & $10.0 \pm 0.1$ &  $10.5 \pm 1.4$  & $0.44$                     & (13) \\
OGLE-2016-BLG-1469                      &  $0.05 \pm 0.01$   & $13.6 \pm 2.1$   & $4.47 \pm 0.51$  & $99.7 \pm 0.8$     &  $0.9 \pm 0.1$    & $0.43$                      & (14) \\
OGLE-2016-BLG-1266$^{\rm (d)}$ &  $0.016 \pm 0.002$ & $11.8 \pm 0.7$ & $3.03 \pm 0.19$ & $8.68 \pm 0.08$   & $9.4 \pm 0.5$   & $1.01 \pm 0.09$          & (15)
\enddata
\tablecomments{ $M_{\rm host}$ and $M_{\rm comp}$ are the masses of the host and companion, respectively.
Except (a), (b), (c), and (d), the masses of 13 binary lens systems are obtained from ground-based parallax measurements.
For the (a), the lens mass is obtained based on the fluxes and colors of the source, while for the (b) and (c), it is obtained from space-based parallax measurements.
$\dagger$ For errors of $\pie$, only values provided in each paper are presented. This is because for cases of providing only errors of $\pien$ and $\piee$, it is not clear that the correlation between $\pien$ and $\piee$ is linear or not. If the correlation is not linear like the event MOA-2016-BLG-231, it is difficult to estimate the error of $\pie$.\\
\textbf{References.} (1) \citet{park2013}, (2) \citet{ranc2015}, (3) \citet{choi2013}, (4) \citet{street2013}, (5) \citet{shin2012}, (6) \citet{han2013}, (7) \citet{jung2015}, (8) \citet{park2015}, (9) \citet{han2016}, (10) \citet{han2017a}, (11) \citet{shvartzvald2016}, (12) \citet{ryu2017}, (13) \citet{shvartzvald2017}, (14) \citet{han2017b}, and (15) \citet{albrow2018}.}
\end{deluxetable}

\begin{deluxetable}{lcccccc}
\tablewidth{0pt}
\tablecaption{Lensing parameters.\label{tbl-two}}
\tablehead{
   \colhead{}                      &  \multicolumn{2}{c}{Best-fit solutions}                            &   \multicolumn{4}{c}{Four models with $\delcs \sim 9$ from Figs. 4 \& 5.}  \\
\cline{2-7} \\ [-1.0ex]
\colhead{Parameter}       &\colhead{$(+,-)$}                &  \colhead{$(-,-)$}                & \colhead{model 1}    &  \colhead{model 2} &  \colhead{model 3}  &  \colhead{model 4}  
 }
\startdata
$\chi2$/dof                     &   $1274.14/1261$                &  $1276.06/1261$                   &   $1282.396/1261$ &   $1282.399/1261$ &    $1282.951/1261$ &    $1283.129/1261$\\
$t_0$ (HJD')                    & $7534.6227 \pm 0.0268$ &  $7534.6419 \pm 0.0249$  &  $7534.6442$        &   $7534.6031$        &    $7534.6157$       &    $7534.6284$\\
$u_0$                               &  $0.1976 \pm 0.0037$       &  $-0.1960 \pm 0.0036$       &   $0.1934$              &   $0.1995$              &    $0.1987$              &    $0.1947$\\
$\te$ (days)                    &  $12.9168 \pm 0.1006$      &  $12.9902 \pm 0.1188$        &  $12.9337$             &   $13.0267$            &   $13.0604$             &    $12.9333$\\  
$s$                                   &  $1.3251 \pm 0.0040$       &  $1.3260 \pm 0.0042$         &  $1.3235$               &    $1.3294$             &    $1.3294$              &    $1.3242$\\
$q$                                   &  $0.4258\pm 0.0080$       &  $0.4163 \pm 0.0080$        &  $0.4167$               &   $0.4251$             &    $0.4218$              &    $0.4219$\\
$\alpha$ (rad)                  &  $4.2705 \pm 0.0036$      &  $-4.2741 \pm 0.0035$       &  $4.2745$              &   $4.2683$            &     $4.2724$             &    $4.2726$\\
$\rho$                               &  $0.0312 \pm 0.0005$      &  $0.0306 \pm 0.0005$        &   $0.0309$             &    $0.0309$            &     $0.0307$            &    $0.0313$\\
$\pien$                              &  $-0.7598 \pm 0.3156$    &  $-0.4996 \pm 0.3047$      &   $-0.0896$           &     $-0.8265$          &     $-1.0557$           &    $-1.5830$\\
$\piee$                              &  $-0.2283 \pm 0.1370$     &  $-0.1749 \pm 0.1219$       &   $-0.1325$            &     $-0.1777$           &     $-0.4287$          &    $-0.6839$\\
$ds/dt$ (yr$^{-1}$)          &  $-0.4685 \pm 0.2488$   &  $-0.2251 \pm 0.2943$       &$-0.5464$              &     $-0.3029$          &     $-0.2988$          &    $-0.6677$\\
$d\alpha/dt$ (yr$^{-1}$) & $-0.5675 \pm 0.3229$     &  $1.2918 \pm 0.3137$          & $-1.1243$              &     $-0.4856$          &     $-0.6642$          &    $-0.5712$\\
$f_{s,\rm spitz}$              &  $53.0998 \pm 1.1631$    &  $51.6615 \pm 1.5408 $        &  $52.7683$            &     $52.3067$         &      $50.3992$         &    $52.9877$\\
$f_{b,\rm spitz}$              &  $-4.5176 \pm 1.1751$    &  $-2.9948 \pm  1.5462$         &  $-4.1389$            &   $-3.5522$          &      $-2.1597$          &   $-4.5540$\\
$f_{s,\rm ogle}$               &  $9.0629 \pm 0.1684$    &  $8.8468 \pm 0.1975 $           &  $8.9815$              &    $8.9539$           &      $8.8901$            &    $9.0908$\\
$f_{b,\rm ogle}$               &  $0.3316 \pm 0.1677$    &  $0.5470 \pm  0.1966$            &  $0.4127$              &     $0.4393$           &      $0.5024$            &    $0.3020$
\enddata
\tablecomments{HJD$^\prime$ = HJD - 2450000.}
\end{deluxetable}

\begin{deluxetable}{lcccccc}
\tablewidth{0pt}
\tablecaption{Physical lens parameters. \label{tbl-three}}
\tablehead{
\colhead{Parameter}                                &        \colhead{$(+,-)$}                     &    \colhead{$(-,-)$}                      &  \colhead{model 1} & \colhead{model 2} & \colhead{model 3} & \colhead{model 4} 
 }
\startdata
$M_{\rm tot}$ $(M_\odot)$                     &    $0.029^{+0.016}_{-0.007}$    &   $0.037^{+0.028}_{-0.011}$    &  $0.180$    & $0.034$    &   $0.025$   &  $0.016$\\ 
$M_{1}$ $(M_\odot)$                              &    $0.020^{+0.011}_{-0.005}$      &   $0.026^{+0.020}_{-0.008}$  &  $0.127$    & $0.024$    &   $0.018$    &  $0.012$\\
$M_{2}$ $(M_\odot)$                              &    $0.009^{+0.005}_{-0.002}$    &   $0.011^{+0.008}_{-0.003}$   &  $0.053$    & $0.010$    &   $0.008$   &  $0.005$\\
$a_\perp$ (au)                                          &    $0.88^{+0.27}_{-0.15}$           &    $1.03^{+0.36}_{-0.21}$         &  $1.96$      & $0.98$      &  $0.80$      &  $0.59$\\
$D_{\rm L}$ (kpc)                                     &    $2.85^{+0.88}_{-0.5}$            &    $3.33^{+1.16}_{-0.67}$         &  $6.33$      & $3.14$      &  $2.57$      &   $1.93$\\ 
(KE/PE)$_{\perp}$                                    &    $0.16^{+0.20}_{-0.10}$           &    $0.52^{+0.30}_{-0.21}$         &  $0.76$      & $0.10$      &   $0.13$      &   $0.09$\\
$\mu_{\rm L,hel,N}\ \rm(mas\ yr^{-1})$  &  $-6.16^{+0.40}_{-0.38}$          &    $-6.09^{+0.43}_{-0.39}$       &  $-3.67$    & $-6.35$    &   $-6.01$    &   $-5.87$\\
$\mu_{\rm L,hel,E}\ \rm(mas\ yr^{-1})$   &  $-0.37^{+0.25}_{-0.39}$         &    $-0.87^{+0.34}_{-0.74}$       &  $-5.03$     & $-0.03$    &   $-0.72$    &   $-0.10$\\
$v_{\rm L,pec,l}\ \rm(km\ s^{-1})$           & $-144.12^{+37.69}_{-50.40}$  &  $-172.29^{+38.72}_{-66.15} $ &  $-331.79$ & $-160.13$ &  $-129.25$ &  $-90.08$\\
$v_{\rm L,pec,b}\ \rm(km\ s^{-1})$          & $-27.78^{+5.95}_{-7.79}$        &  $ -24.98^{+7.17}_{-7.71}$        &  $ 85.82$   & $ -38.37$  & $-20.73$   &  $-18.29$
\enddata
\tablecomments{Physical lens parameters for $(+,-)$ and $(-,-)$ solutions represent the median values of each physical parameter from the $(+,-)$ and $(-,-)$ MCMC chains, and their error bars represent the 16th and 84th percentile values of each physical parameter from the chains. This is because the correlation between $\pien$ and $\piee$ components is not well approximated by a linear relation, as shown in Figure 4.}
\end{deluxetable}

\begin{figure*}
\centering
\includegraphics[width=150mm]{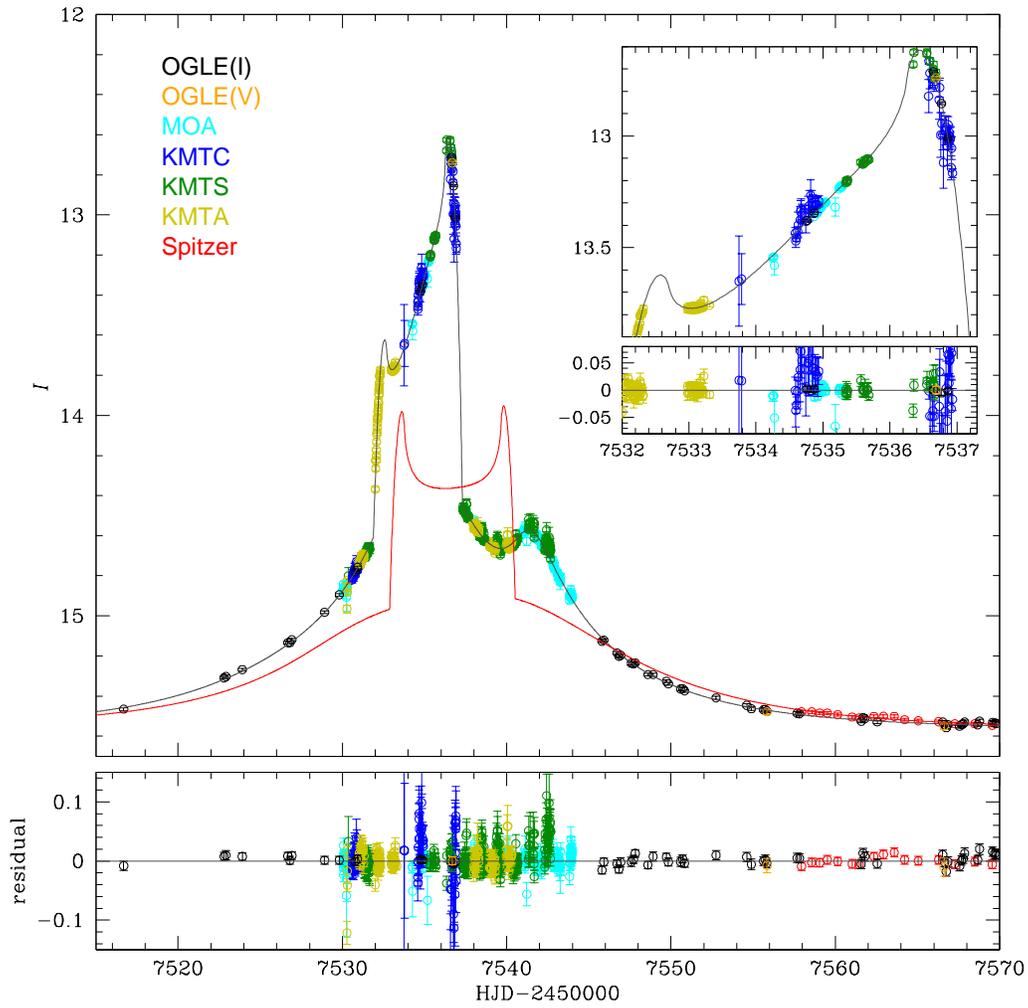}
\caption{\label{fig:one}
Light curves of the best-fit binary model, $(+,-)$ model, for MOA-2016-BLG-231, including both microlens parallax and orbital effects.
Black and red lines represent the light curves as seen from Earth and \textit{Spitzer}, respectively.
}
\end{figure*}

\begin{figure*}
\centering
\includegraphics[width=150mm]{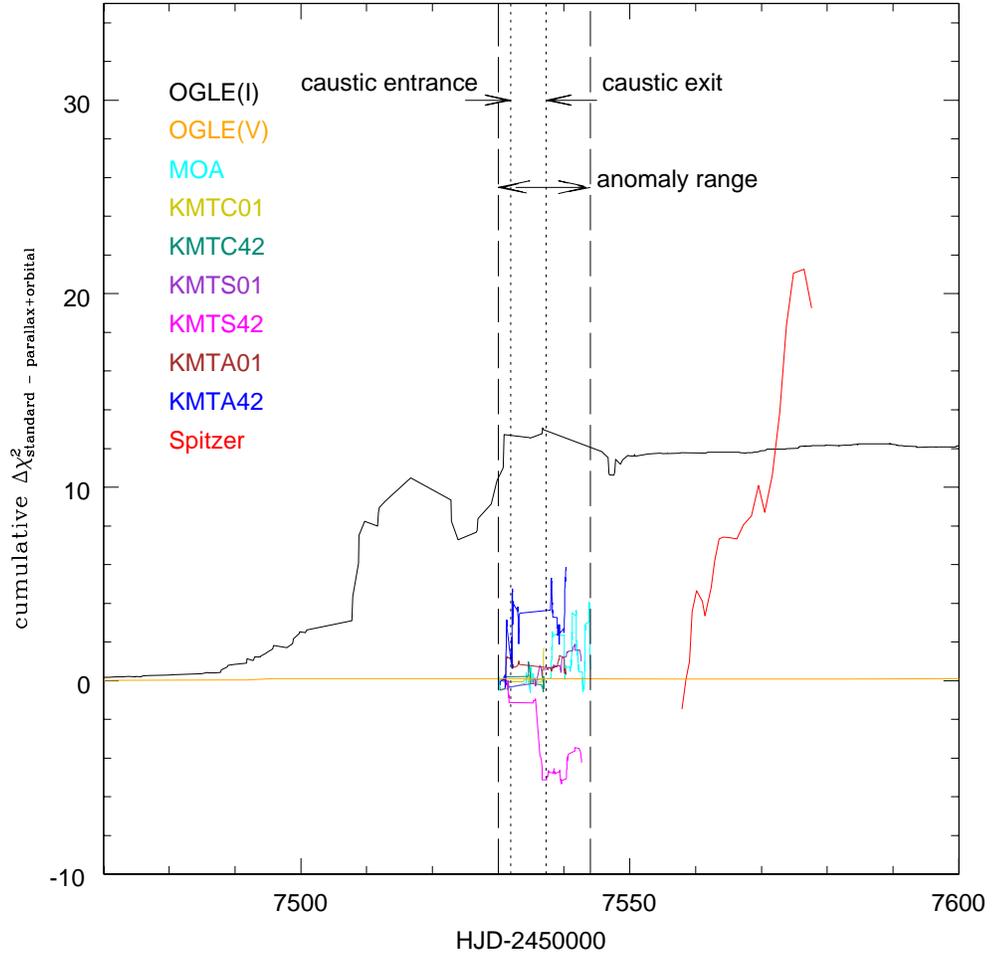}
\caption{\label{fig:two}
Cumulative distribution of $\delcs$ between the standard and the parallax+orbital models for $(+,-)$.
All ground-based data except OGLE are used only in the anomaly range $7530.0 < {\rm HJD - 2450000\  (HJD}^\prime) < 7544.0$. 
The epochs of the entrance and exit of the caustic are $\rm HJD^\prime =7531.9\ {\rm and}\ 7537.3$, respectively.
}
\end{figure*}

\begin{figure*}
\centering
\includegraphics[width=150mm]{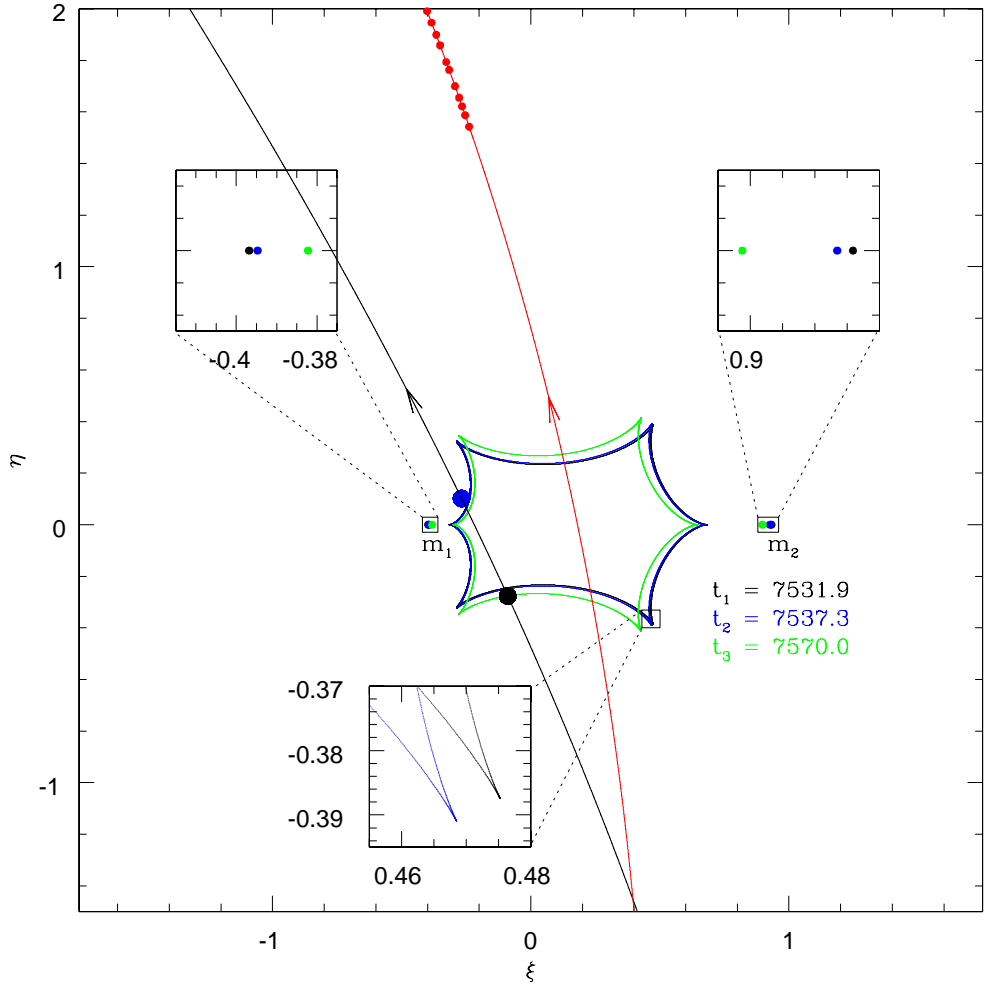}
\caption{\label{fig:three}
Source trajectories for the best-fit $(+,-)$ model as seen from the ground (black) and \textit{Spitzer} (red), with the red points indicating the epochs of \textit{Spitzer} data.
The caustic structure and positions of the binary lens components change with time due to the lens orbital motion, and these changes are shown at three epochs, $t_1 = 7531.9$ (caustic entrance), $t_2 = 7537.3$ (caustic exit), and $t_3 = 7570.0$ (close to baseline).
However, the caustic and lens components at the two epochs $t_1$ and $t_2$ overlap and so require zooms to see these effects (see insets).
The black and blue solid circles on the caustic curve represent the source positions at $t_1$ and $t_2$, respectively.
$M_1$ and $M_2$ are the primary and secondary components of the binary.
}
\end{figure*}

\begin{figure*}
\centering
\includegraphics[width=150mm]{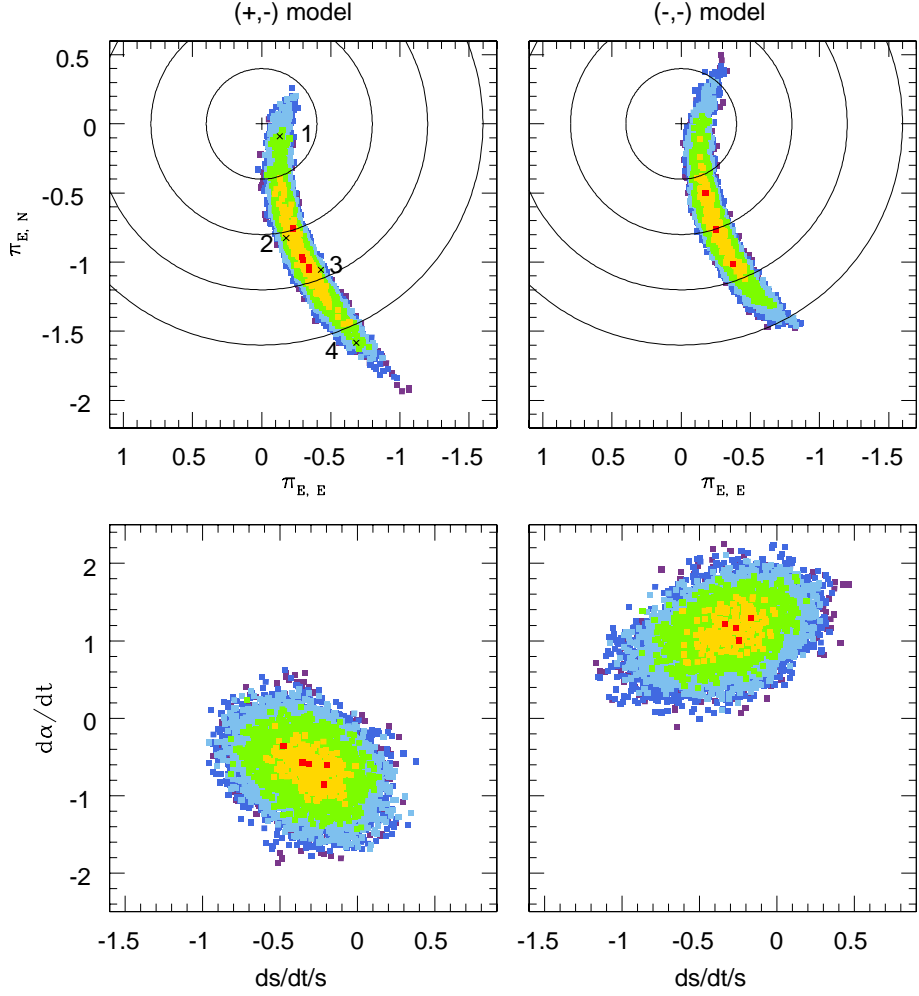}
\caption{\label{fig:four}
$\delcs$ distributions of the microlens parallax and orbital motion parameters for the two best-fit models obtained from the parallax+orbital modeling.
The red, yellow, green, light blue, dark blue, and purple colors represent regions with $\delcs < (1, 4, 9, 16, 25, 36)$, from the best-fit model, respectively.
The four solid circles in the parallax distributions are centered at $(\pien,\piee) = (0,0)$ and have the radii of $\pie = (0.4, 0.8, 1.2, 1.6)$, respectively.
Four parallax models, which are located inside the $3\sigma$ contour, are marked as 1, 2, 3, and 4 in the distribution of the $(+,-)$ model.
The trajectories and light curves for these four models are illustrated in Figure 5.
}
\end{figure*}

\begin{figure*}
\centering
\includegraphics[width=150mm]{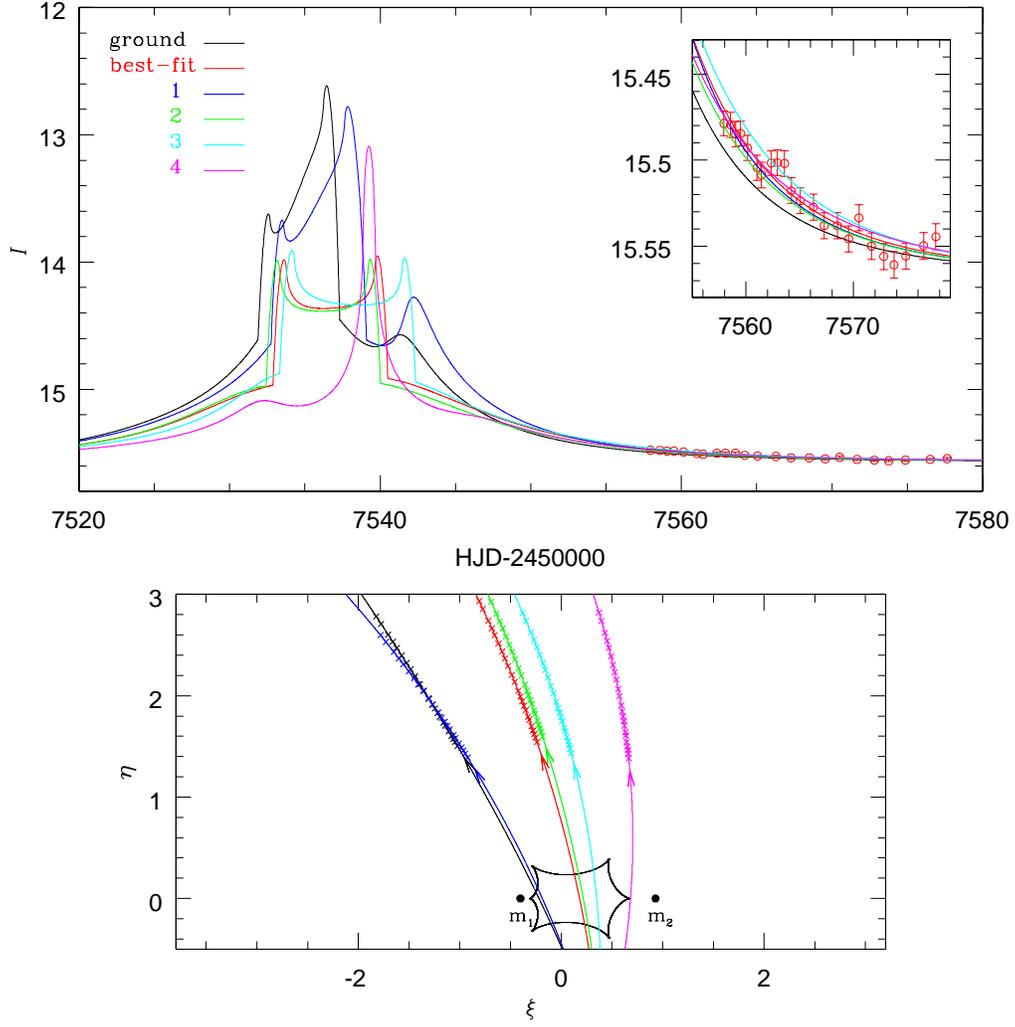}
\caption{\label{fig:five}
Upper panel : \textit{Spitzer} light curves for the best-fit model and the four $\delcs \sim 9$ models highlighted in Figure 4.
Each model is drawn with different colors.
The black curve represents the best-fit ground-based light curve.
The \textit{Spitzer} data are plotted on the best-fit light curve.
Bottom panel : Corresponding \textit{Spitzer} and ground-based trajectories.
They are drawn with the same color as in the upper panel.
}
\end{figure*}

\begin{figure*}
\centering
\includegraphics[width=150mm]{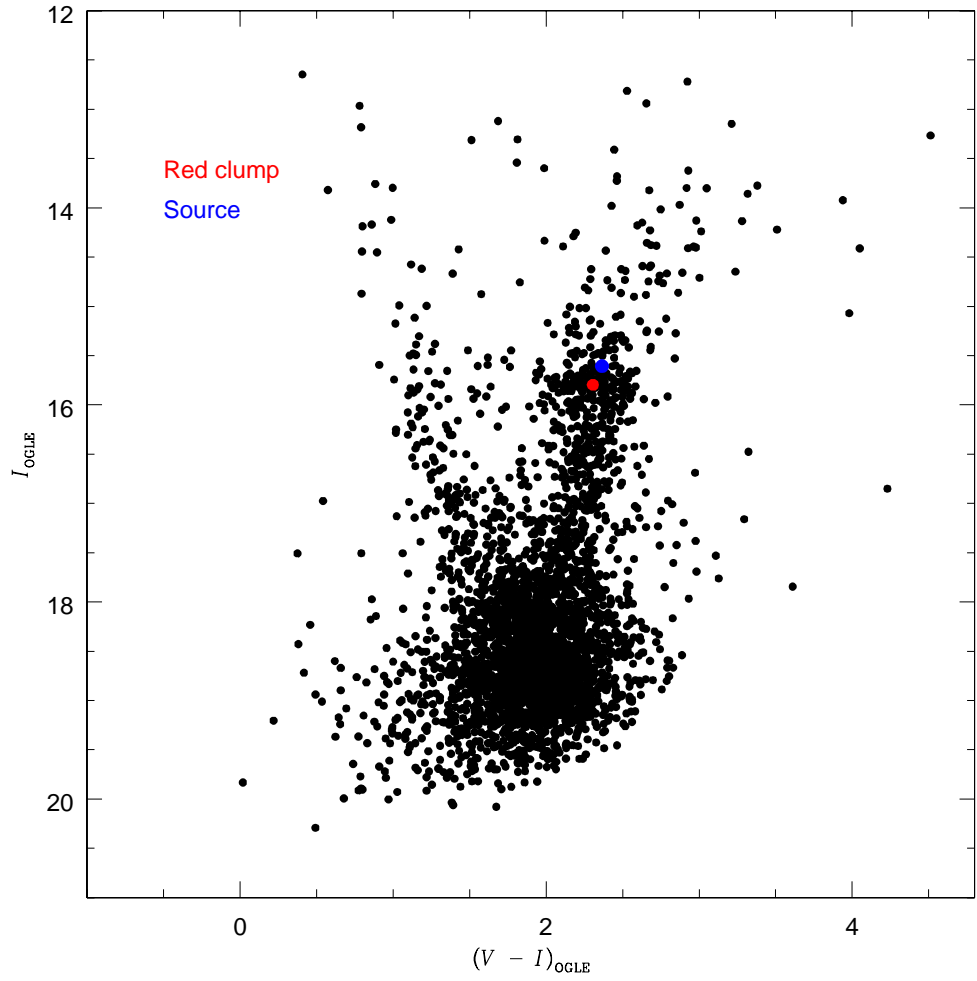}
\caption{\label{fig:six}
Instrumental color-magnitude diagram (CMD) of stars in the observed field.
The red and blue circles mark the centroid of the red clump giant and microlensed source star, respectively.
}
\end{figure*}

\begin{figure*}
\centering
\includegraphics[width=150mm]{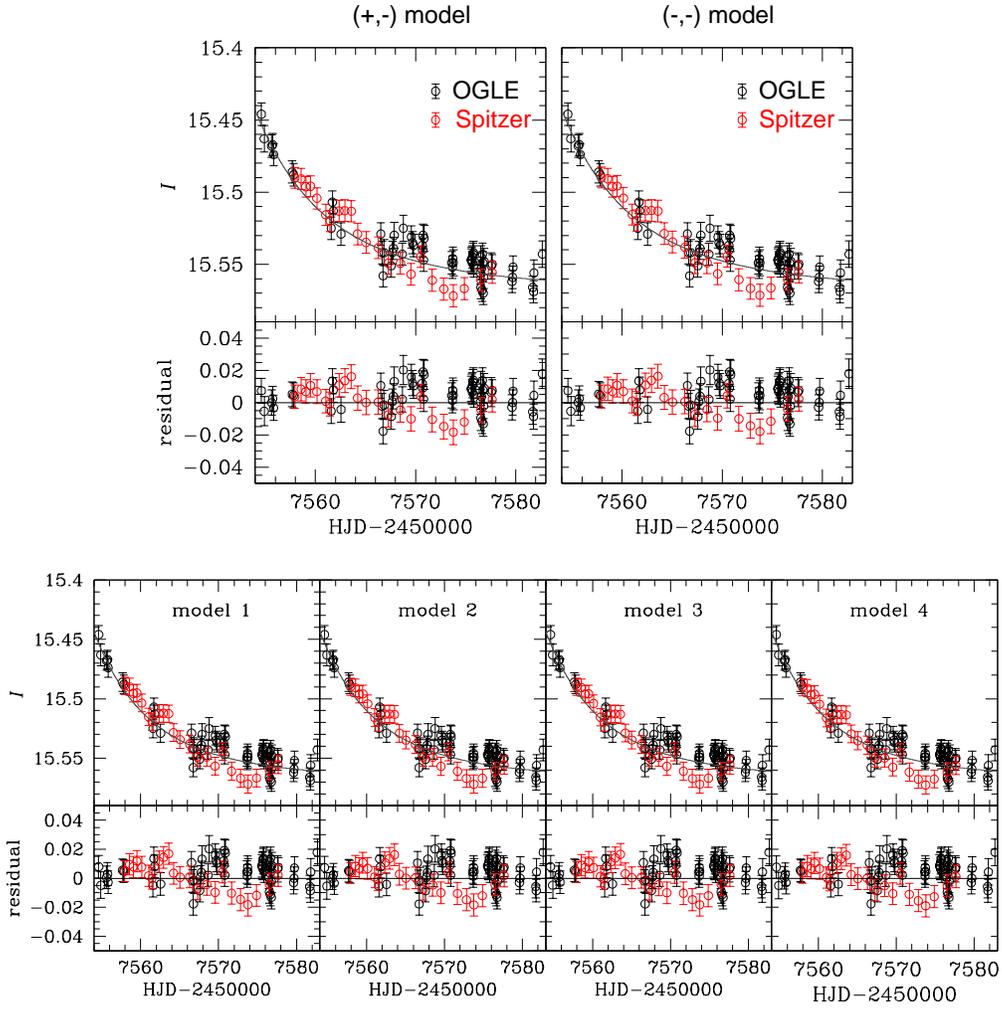}
\caption{\label{fig:seven}
Light curves for the standard model, which is conducted under the conditions that the high-order effect parameters are set to zero and the initial values of the other standard model parameters including \textit{Spitzer} fluxes are set to the best-fit solutions of $(+,-)$ and $(-,-)$ models (Upper panel) and the solutions of four models indicated in Figure 4 (Bottom panel).
}
\end{figure*}

\begin{figure*}
\centering
\includegraphics[width=150mm]{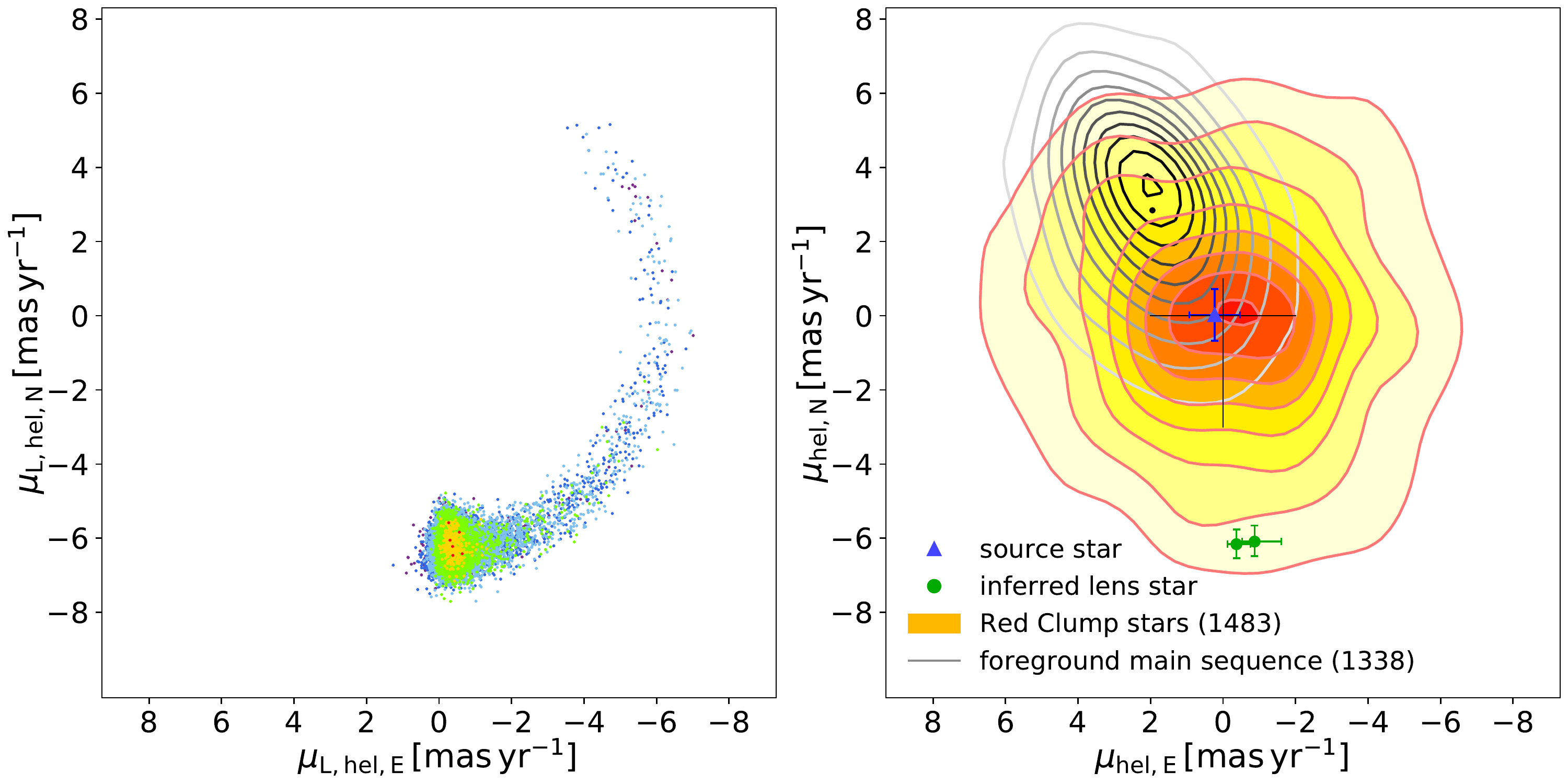}
\caption{\label{fig:eight}
Left panel: Scatter of the heliocentric proper motion of the lens for the $(+,-)$ best-fit model. The color notation is the same as Figure 4.
Right panel: Proper motion of stars in the observed field within a $6.5'\  \times\  6.5'$ square.
Orange and grey contours represent the proper motions of red clump stars and main sequence stars, which correspond to the bulge and disk populations, respectively.
The blue triangle is the proper motion of the source, while the green dots are the lens proper motions of two degenerate solutions, $(+,-)$ and $(-,-)$.
The source is essentially at rest with respect to the bulge stars.
}
\end{figure*}

\end{document}